\documentclass[11pt,longbibliography]{revtex4-1}

\usepackage{amsmath}    
\usepackage{mathenv}
\usepackage{graphicx}   
\usepackage[caption=false]{subfig}
\begin{document}

\title{Inductance extraction of superconductor structures with internal current sources}

\author{M.~M.~Khapaev}
\affiliation{Dep. of Computer Science, Lomonosov Moscow State University, Moscow, Russia}
\author{M.~Yu.~Kupriyanov}
\affiliation{Skobeltsyn Institute of Nuclear Physics, Lomonosov Moscow State University, Moscow, Russia}
\affiliation{Solid State Physics Department, Kazan Federal University, Kazan, Russia}

\date{\today}

\begin{abstract}
The sheet current model underlying the software 3D-MLSI package for calculation of inductances of multilayer superconducting circuits, has been further elaborated. The developed approach permits to overcome serious limitations on the shape of the circuits layout and opens the way for simulation of internal contacts or vias between layers. Two models for internal contacts have been considered. They are a hole as a current terminal and distributed current source. Advantages of the developed approach are illustrated by calculating the spatial distribution of the superconducting current in several typical layouts of superconducting circuits. New meshing procedure permits now to implement triangulation for joint projection of all nets thus improving discrete physical model for inductance calculations of circuits made both in planarized and non-planarized  fabrication processes. To speed-up triangulation and build mesh of better quality, we adopt known program "Triangle".
\end{abstract}

\maketitle

\section{Introduction}

The challenges \cite{Anders,IARPA} facing the development of the modern digital superconducting electronics urgently require not only the development of new technological solutions \cite{Tolpygo2014,Tolpygo22014,Nagasawa,Fujima,Nagasawa2} but also new tools needed to calculate  inductances, resulting in topological configurations of designed digital cells. Inductances are the  important component of all superconductor digital circuits.
Calculation of inductances, currents and fields for layouts in superconductor electronics is important and challenging problem \cite{GajFeldman,Fourie2013Status}. Currently several programs are used for inductances calculations \cite{Lmeter,CoenradFourier2011,khapaev20013d}. These programs are intended for different areas \cite{Fourie2013Status} and utilize different superconducting current models. Recently it was demonstrated \cite{2014arXiv1408.5828T}  that 3-D inductance extractors based on FastHenry \cite{InducEx,FASHENRY0,FastHenry3.0wr}
%
%
and 3D-MLSI \cite{khapaev2001inductance, kupriyanov2010}
software can be successfully used for calculations of inductance of various superconducting microstrip-line and stripline inductors having linewidth down to 250 nm in 8-metal layer process developed for fabricating VLSI superconductor circuits.


Unfortunately, the existing inductances extraction tools have some limitations. Lmeter \cite{Lmeter} do not apply, if parts of a film or a wire in a multilayer structure
don't have strong magnetic coupling with other layers in the structure. For example, Lmeter can't be used for single layer structures and structures without groundplanes.
FastHenry tool \cite{InductExCalibration} needs accuracy calibration and meets problems for holes and groundplanes. It is difficult to use 3D-MLSI \cite{khapaev2001inductance} for quantitative and qualitative description of the effects caused by current injection through the internal terminals located inside multilayer structures. These terminals are staggered or stacked vias between layers \cite{2014arXiv1408.5828T} or connections between the films contained  Josephson junction.


In this paper we attack these problems by improvement of our 3D-MLSI software aimed on removing 
limitations on using the internal terminals. To do that we introduce two new models for current sources and improve the accuracy of our numerical algorithm and program by using the new scheme of FEM triangular meshing aligned to all film boundaries. The scheme allows to do more accurate calculations for non-planarized circuits and has as an option allowing us to use an external program Triangle \cite{shewchuk96b} for FEM mesh construction.

In the first model of internal terminal we declare a hole or any part of hole in a multilayer film as current terminal and define inlet or outlet current on its perimeter. This new current terminal is almost identical to a similar terminal located at the external borders of the film.
However, there is the difference. It consists in the fact that
the new mutual inductance between current around the hole and current from hole appears.

The first model doesn't allow a current flows under the contact. It isn't applicable if there are two contacts on same place from top and from bottom of the film.
In these cases it is convenient to use the second internal terminal model. We call it "hole as a current source".


In the second model, the area of the film, which is located under the contact (via) is not cut out. It remains an integral part of the film, in which we solve the equations that determine the spatial distribution of the current. These equations are properly modified to include current sources located in the area of inner terminals. 
The total current provided by the current source is the given value.

%

Advantages of the developed approach are illustrated in the last sections of the paper by calculation of the spatial distribution of the superconducting current in several typical layouts of superconducting circuits.

\section{Basic Assumptions}

We consider multilayer, planar, multi-connected structures, which consist of superconducting (S) films separated
by dielectric interlayers.
The design can have or have not ground plane that reside under all wires. 
There are no restrictions on floor plane shapes of the S films. They
 can contain holes, current terminals on external boundary and current terminals (contacts) in inner areas of S layers.
Current distribution in the film can be induced by different sources. These sources can be given full currents circulating around holes or fluxoids trapped in the holes, given full currents between external or internal contacts, and external magnetic field. 

\begin{figure}[ht]
\begin{center}
\includegraphics{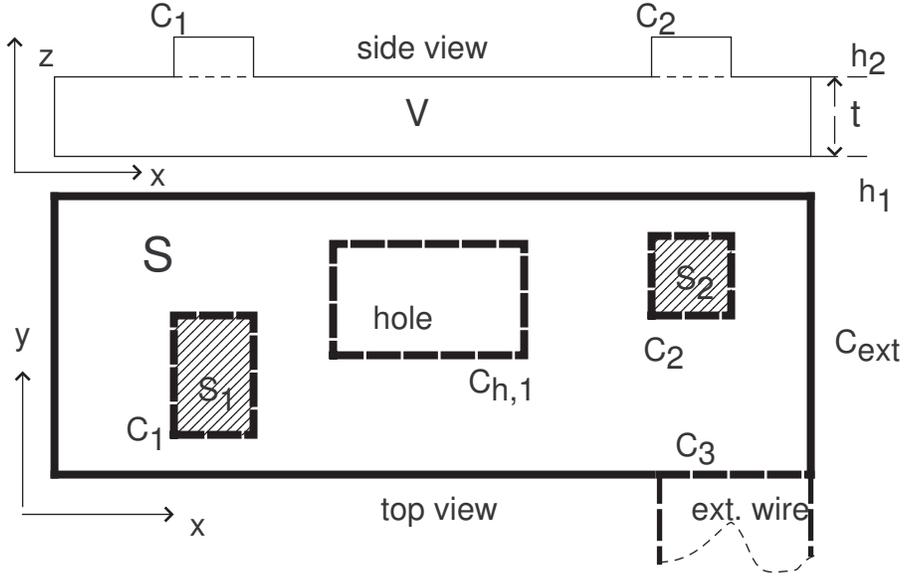}
\end{center}
\caption{Schematic view of a single layer circuit design having two internal contacts (dashed area), one contact on external boundary and hole.
$C_1$ and $C_2$ are boundaries of internal contacts.
$C_3$ is contact on external boundary.
$C_{ext}$ is external boundary without contacts.
Currents can flow around holes and from contact to contact.} \label{schem}
\end{figure}

A single S film with one hole and three current terminals (contacts) is shown in Fig. \ref{schem}. It will be used to illustrate new features of the presented version of our 3D-MLSI package.
The hole on Fig. \ref{schem} traps zero or non-zero flux. Internal contacts can model Josephson junctions, as well as staggered or stacked vias between S layers. Terminal on external boundary (dashed segment) models external wire. 


For further convenience, let $P$, $P'$ stands for points in 3D space, $r$, $r'$ - for points on 2D plane. Also, consider
differential operators $\partial_x={\partial}/{\partial x}$, $\partial_y={\partial}/{\partial y}$,
$\nabla=(\partial_x,\partial_y,\partial_z)$, $\nabla_{xy}=(\partial_x,\partial_y)$. $\Delta$ is Laplace operator in 3D and $\Delta_{xy}$ is Laplace operators in 2D space.

\section{Mathematical Model}

Rigorous electromagnetic analysis should be started from stationary Maxwell and London equations \cite{Orlando:book,van1998principles}:
\begin{align}
\lambda^2\,\nabla\times\vec{j}(P)+ \vec{H}(P)+\vec{H}_{ext}(P) = 0,\label{ell} \\
\nabla\times\vec{H}(P)=\vec{j}(P),\label{emx}
\end{align}
where $\lambda$ is London penetration depth, $\vec{H}(P)$ is magnetic field of current $\vec{j}(P),$ $\vec{H}_{ext}(P)$ is external magnetic field.
Equations (\ref{ell}), (\ref{emx})
can be rewritten in the form of volume current integral equations using vector potential, $\vec{A}_{ext}(P)$ for magnetic field
\begin{align}
\lambda^2\vec{j}(P)+\frac{1}{4\pi}\int\!\!\!\!\!\int\!\!\!\!\!\int\limits_{V}\frac{\vec{j}(P')}{|P-P'|}dv'+\vec{A}_{ext}(P)=
\nabla\chi(P), \quad r\in V \label{ev1} \\
\quad \nabla\cdot\vec{j}(P)=0,\quad \Delta\chi(P)=0, \quad  \vec{H}_{ext}(P) = \nabla\times\vec{A}_{ext}(P).\label{ev2}
\end{align}
Here integration is performed over the volume $V$ of all conductors,
$\chi(P)$ is a scalar function, which is proportional to phase of superconductor condensate function.

Equations (\ref{ev1}), (\ref{ev2}) together with appropriate boundary conditions can be solved numerically.  Typically, to do that the PEEC (Partial Element Equivalent Circuit) method is used. This method was evaluated for normal conductors \cite{peecRuehli}, enhanced for large problems \cite{FASHENRY0} and recently adopted for superconductors \cite{FastHenry3.0wr,CoenradFourier2011}.

Boundary conditions for Eqs. (\ref{ev1}), (\ref{ev2}) are easy formulated for wire-like conductors with external or internal current terminals. Description of holes with trapped fluxoids and large flat structures like ground planes meets some difficulties in using these PEEC-like methods. For superconductors it can cause accuracy, memory and performance problems.

Our approach is based on some assumptions concerning dimensions of the circuits. 
We assume that floor plan dimensions are much larger than the film thicknesses that in turn are less or of the order of London penetration depth. In this case, the volume current density in superconductor can be accurately approximated by a sheet current density. If the assumptions are violated then the accuracy of our approach is reduced.
Nevertheless, the method provides sufficient accuracy of calculation in the case where the film thickness is about 2 - 3 penetration depths \cite{khapaev2001inductance}.

Planarity assumptions allow us to introduce sheet current $J(r)$. Let $t=h_2-h_1$ is the thickness of the layer (see Fig. \ref{schem}) and $\lambda_S=\lambda^2/t$ is London penetration depth for films. We assume that $j(P)=(j_x(P),j_y(P),0)$ and take average volume current density over the film thickness (Fig. \ref{schem}):
\begin{align}
\vec{J}(r)=\frac{1}{t}\int_{h_1}^{h_2} \vec{j}(P)dz.\label{avgj}
\end{align}
Then from (\ref{ev1}) it follows that $\vec{J}(r)$
satisfies the integral equation:
\begin{align}
\lambda_S\vec{J}(r)+\frac{1}{4\pi}\int\!\!\!\!\int\limits_{S}G(r,r')\vec{J}(r')ds'=
\nabla_{xy}\chi(r), \label{esh1} \\
\quad \nabla_{xy}\cdot\vec{J}(r)=0,\quad \Delta_{xy}\chi(r)=0. \quad r\in S.
\end{align}
Kernel $G(r,r')$ is result of averaging procedure for (\ref{ev1}). For single layer problems it can be taken simply as
\begin{align}
G(r,r')=\frac{1}{|r-r'|}. \label{G_simple}
\end{align}
For multilayer structures with layers $m$ and $n$ and heights $h_{m,k},h_{n,l}$ \cite{khapaev2001inductance}
\begin{align}
G_{mn}(r,r')=\frac{1}{4}\sum_{k=1}^2\sum_{l=1}^2\left(|r-r'|^2+(h_{m,k}-h_{n,l})^2\right)^{-1/2}.
\label{Gmn}
\end{align}

On the next step it is convenient to introduce the  {\em stream function} $\psi(r)$
%
\begin{align}
J_{x}(r) = \partial_y \psi(r)
, \quad
J_{y}(r) =-\partial_x \psi(r)
\label{pssi}
\end{align}
and rewrite (\ref{esh1}) in the form \cite{khapaev2001inductance}
%
\begin{align}
-\lambda_S\Delta_{xy}\psi(r)+\frac{1}{4\pi}\int\!\!\!\!\int\limits_{S}\!\! \left(
\nabla_{xy}\psi(r'),\nabla_{xy}'G(r,r')\right)ds_r'+H_{z,ext}(r)=0.
\label{mp}
\end{align}
Here $H_{z,ext}(r)$ is component of external magnetic field oriented in $z$ direction. For very thin conductors $G(r,r')$ can be taken in the form  (\ref{G_simple}).

Equation (\ref{mp}) should be supplemented by boundary conditions. These boundary conditions are simple first kind boundary conditions since  values of stream function on the boundary are known \cite{khapaev2001inductance}:
\begin{align}
\psi(r)=I_{h,k},\; r\in C_{h,k},\;
\label{holes} \\
\psi(r)=F(r),\; r\in C_{ext}.
\label{terms}
\end{align}
Here $I_{h,k}$ is the full current circulating around hole $k$ with boundary $C_{h,k}$. On the external boundary $C_{ext}$ function  $F(r)$ can be easily evaluated using well-known properties of stream function.

Mathematically problem (\ref{mp}), (\ref{holes}), (\ref{terms}) is very similar to boundary problem for Poisson equation.
We prefer to solve equation (\ref{mp}) instead of (\ref{esh1}) since (\ref{mp}) easily accounts currents circulating around holes and
for reasons of efficiency of numerical computations.
After calculation of the $\psi(r)$ function, we can calculate the energy functional, as well as the inductance matrix \cite{khapaev2001inductance}.

Unfortunately $\psi$-function approach meets problems for structures with internal contacts as contacts 1 and 2 in Fig. \ref{schem}. This problem is a purely mathematical \cite{ZRen2003}. It isn't possible to define stream function for internal source.
There are artificial approaches to resolve this problem, which are based on the introduction of the cuts
between contours of internal terminals and external boundary.
But it is just workaround and not a practical solution.

To overcome these difficulties we decompose current density into the sum of excitation current for terminals and screening current:
\begin{align}
\vec{J}(r)=\vec{J}_{ex}(r)+\vec{J}_{scr}(r). \label{deco}
\end{align}
For evaluating excitation current $\vec{J}_{ex}(r)$, some techniques are known \cite{ZRen2003,Tamburrino2010}. These techniques are based on topological considerations for finite element method meshes and as result produce non-physical currents for so called "thick cuts" for internal sources \cite{ZRen2003}. In our case it is still difficult to account all full current combinations for calculations of elements of inductance matrix.

Fortunately one more physical decomposition (\ref{deco}) exists. Physically, it is equivalent to  the separation of the total current on the  circulating and laminar components.
To implement it, taking into account (\ref{esh1}), we define excitation current as
\begin{align}
\vec{J}_{ex}(r)=\frac{1}{\lambda_S}\nabla_{xy}\varphi(r), \quad
\Delta_{xy}\varphi(r)=0,  \quad
\frac{1}{\lambda_S}\frac{\partial\varphi}{\partial n}=0,\quad r\in C_{ext},C_{h,k}. \label{Jex}
\end{align}

Equation (\ref{Jex}) needs boundary conditions for internal sources and contacts on the external boundary. We consider two approaches for internal sources modeling.

In the first model we consider internal contacts as holes. In this case we have two current components. One is flowing across hole boundary and the other is circulating around the hole. Current across boundary for internal and external sources should be presented by Neumann boundary conditions for function $\varphi(r)$:
\begin{align}
\frac{1}{\lambda_S}\frac{\partial\varphi}{\partial n}=\frac{I_m}{|C_m|},\quad r\in C_m. \label{bch}
\end{align}
Here $C_m$ is the boundary of $m$-th contact, $I_m$ is full current across $C_m$ and $|C_m|$ is the 
length of contact. It is assumed that the injection current is distributed uniformly along the perimeter of any internal or external terminals. This assumption is physically justified since in real devices the characteristic dimensions of the terminal is much smaller than $\lambda_{S}$.

The first model
allows us to investigate new objects such as mutual inductance of hole and contact to this
hole.

The first model has two disadvantages.
The current can't flow under contact. Also contacts from top and bottom on same place are a problem. In this case first approach brings to two different intersecting holes.
Both of these drawbacks are overcame in the second model of the internal terminal.

In the second model, the terminal is considered as the locus of local current sources $\vec{J}_{ex}(r)$.
\begin{align}
\nabla_{xy} \cdot \vec{J}_{ex}(r)=\frac{I_m}{|S_m|},\quad r\in S_m, \label{div}
\end{align}
where $I_m$ is the full current injected into the area, $|S_m|,$ of internal contact $m.$

For structure in Fig. \ref{schem} it brings us to the following boundary problem for function $\varphi(r)$:
\begin{align}
\vec{J}_{ex}(r)=\frac{1}{\lambda_S}\nabla_{xy}\varphi(r), \label{Jdiv1} \\ 
\frac{1}{\lambda_S}\Delta_{xy}\varphi(r)=F(r),  \quad F(r)=\frac{I_m}{|S_m|} \quad r\in S_m, \quad F(r)=0 \; r\notin S_m, \label{Jdiv2} \\ 
\frac{1}{\lambda_S}\frac{\partial\varphi}{\partial n}=0,\quad r\in C_{ext},C_{h,k},\quad \frac{1}{\lambda_S}\frac{\partial\varphi}{\partial n}=\frac{I_3}{|C_3|},\quad r\in C_3. \label{Jdiv3}
\end{align}
From (\ref{Jex}) it follows that circulation of $\vec{J}_{ex}(r)$ around any hole equals to zero. For $\vec{J}_{scr}(r)$ from (\ref{esh1}) we have
\begin{eqnarray}
-\lambda_S \nabla_{xy} \times \vec{J}_{scr}(r)+ \nonumber \\ 
\frac{1}{4\pi}\int\!\!\!\!\int\limits_{S}\!\!
\left(J_{scr,x}(r')\partial_y G(r,r')-J_{scr,y}(r')\partial_x G(r,r')\right)ds'+F(r)=0 \label{Jscr} \\
\hspace*{-0.5cm}F(r)=\frac{1}{4 \pi \lambda_S}\int\!\!\!\!\int\limits_{S}\!\! \left(\partial_x\varphi(r')\partial_y G(r,r')-\partial_y \varphi \partial_x G(r,r')\right)ds'+H_{ext,z}(r)=0 
\label{Fr}
\end{eqnarray}


After solution of the  boundary problem (\ref{Jdiv1},\ref{Jdiv2},\ref{Jdiv3}) for $\varphi(r)$ it is possible to calculate the $F(r)$ function from  (\ref{Fr}) and reduce the problem to computation of $\vec{J}_{scr}(r)$ making use of the well developed later \cite{khapaev2001inductance, meSST, kupriyanov2010, TEEconf} stream function approach
\begin{align}
J_{scr,x}(r) = \partial_y \psi(r), \quad
J_{scr,y}(r) =-\partial_x \psi(r)
\end{align}
with the boundary conditions for holes
\begin{align}
\psi(r)=0 \quad r\in C_{ext}, \quad, \psi(r)=0 \quad r\in C_{3}, \quad \psi(r)=I_{h,k},\; r\in C_{h,k}.
\label{Jscr_holes}
\end{align}

In accordance with Eq. (14), the vectorial sum of $J_{ex}(r)$ and $J_{scr}(r)$ determines the spatial distribution of the total current $\vec{J}(r)$ in the structure. Knowledge of this distribution allows us to calculate the total energy $E$
\begin{align}
E=\frac{1}{2}\left(\lambda_S \int\!\!\!\!\int\limits_{S}\!\!|\vec{J}(r)|^2+\frac{1}{4\pi}\int\!\!\!\!\int\limits_{S}ds'\int\!\!\!\!\int\limits_{S}
(\vec{J}(r),\vec{J}(r'))G(r,r')ds\right), \label{E}
\end{align}
which in turn makes it possible to  find the inductance matrix \cite{khapaev2001inductance}.

\section{Numerical technique and program}

\subsection{Finite Elements Method}

Our basic numerical technique is Finite Element Method (FEM) \cite{jin2002finite}. We use triangular meshes and linear finite elements. This approach was evaluated for stream function equations (\ref{Jscr}) in \cite{khapaev2001inductance,khapaev20013d,meSST}. For Poisson equations (\ref{Jex}) and (\ref{Jdiv1},\ref{Jdiv2},\ref{Jdiv3}) FEM implementation is strait forward.


There are several CPU time consuming procedures in the algorithm. The first is calculation of FEM approximation of equation (\ref{Jscr}) and the right part $F(r)$ in Eq. (\ref{Fr}). The next is calculation of full energy defined by the expression (\ref{E}). To speed up all of them we introduce matrix of interactions between triangles in FEM mesh.  Let $\Delta_i$ and $\Delta_j$ be two cells - triangles in FEM mesh, then elements of interaction matrix are quadruple integrals
\begin{align}
a_{ij}=\frac{1}{4\pi}\int\!\!\!\!\int\limits_{\Delta_i}ds'\int\!\!\!\!\int\limits_{\Delta_j}G(r,r')ds. \label{aij}
\end{align}
Half-analytic method for (\ref{aij}) evaluation was developed in \cite{khapaev2001inductance,TEEconf} and essentially used in the current version of software. Matrix with elements (\ref{aij}) allows to perform quick and easy calculation of FEM matrixes and the full energy (\ref{E}).

The solution of FEM linear system equations for $J_{scr}$ (\ref{Jscr}) is the third CPU time consuming procedure since it deals with inversion of large fully populated matrix. FEM solutions for (\ref{Jex}) and (\ref{Jdiv1},\ref{Jdiv2},\ref{Jdiv3}) are fast because it based on the use of sparse matrix technique. Nevertheless time of calculations stay acceptable even for rather large problems where FEM matrix dimension can reach $10000$ or more.


To reduce the size of FEM matrices and improve accuracy, we use program Triangle \cite{shewchuk96b} as core engine for FEM mesh construction.  New technique improves meshing for overlapped multilayer structures. For that we implement triangulation for joint projection of all nets. This approach improves discrete physical model for planarized and non-planarized fabrication processes. For non-planarized processes we can assign different film height for every triangle in the mesh. Together with formula (\ref{Gmn}) it gives very accurate discrete model of a layout. All these enhancements allow effective solution of larger problems with smaller FEM meshes and with good accuracy.

\section{Spatial distribution of supercurrent in typical layouts}

\subsection{Hole as terminal}


%
%
%
%
%

\begin{figure}
\centering
\subfloat[]
{
\includegraphics[width=0.5\textwidth] {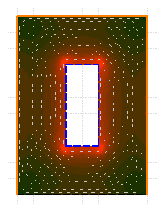}
\label{ht1:1ho}
}
\subfloat[]
{
\includegraphics[width=0.5\textwidth] {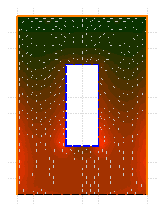}
\label{ht1:1term}
} 
\caption{Spatial distribution of supercurrent in $8\times 11\mu m$ plate with $2\times 5\mu m$ hole for different ways to set the current in the plate via internal terminal. Calculations have done  making use of first internal terminal model. Darker gray corresponds higher current density.  Current direction is shown by vanes. Terminals are shown by dashed lines. Pictures are produced by our program.
a) The current excited by fluxoid $\Phi=L\cdot I$, $I=1ma$, trapped in the hole.
b) The injected current is distributed uniformly around the perimeter of the hole and flows out through the lower border of the plate.
}
\label{ht1:htb}
\end{figure}

\begin{figure}
\centering
\subfloat[]
{
\includegraphics[width=0.5\textwidth] {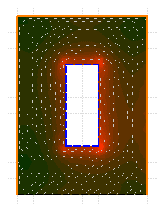}
\label{ht2:1mut}
}
\subfloat[]
{
\includegraphics[width=0.5\textwidth] {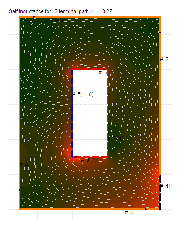}
\label{ht2:side_term}
}
\caption{Spatial distribution of supercurrent in $8\times 11\mu m$ plate with $2\times 5\mu m$ hole for different ways to set the current in the plate via internal terminal. Calculations have done  making use of first internal terminal model. Darker gray corresponds higher current density. Current direction is shown by vanes. Terminals are shown by dashed lines. 
a) Fluxoid trapped in the hole plus current from terminal-hole to terminal-bottom. 
b) The injected current is distributed uniformly on left border  of the hole and flows out through the part of the right border of the plate.
}
\label{ht2:htbb}
\end{figure}


Consider simple inductances calculations demonstrating the model of hole as terminal. The structure we consider is $8\times 11\mu m$ plate with thickness $0.4\mu m$ and with $2\times 5\mu m$ hole and London penetration depth $\lambda = 0.4\mu m$. On Fig. \ref{ht1:htb} current density and current direction with small vanes are shown. When we calculate the inductances we set full currents around holes and full currents flowing across terminals as $I=1 mA$. For hole it means that some fluxoid $\Phi=L\cdot I$ is trapped in the hole where $L$ is the inductance of hole.

Fig. \ref{ht1:1ho} show currents for self inductance of the hole. Simple estimation using per-unit-length inductances of coplanar lines of $3\mu m$ width and $2\mu m$ and $5\mu m$ spacing between lines gives $9.9 pH$. In this estimation we calculate per-unit-length inductances using 2D program \cite{KhapaevJr}. These inductances are $1.03 pH/\mu m$ for spacing $2\mu m$ and $1.24 pH/\mu m$ for spacing $3\mu m$. We take length of strips $6\mu m$ and $3\mu m$ to account corners of the hole.  Our result with 3D-MLSI is $10.1 pH$ and match estimation well.

Fig. \ref{ht1:1term} demonstrates results for hole as terminal. All four sides of the hole inject current with uniform density. Current leave plate across bottom boundary. Inductance of this current path is $1.59pH$.

Also, we can calculate mutual inductance of current circulating around the hole and current from hole to bottom side. This inductance is very small ($3\cdot 10^{-5}$pH ) because the problem is symmetric. Spatial distribution of the current is shown in Fig. \ref{ht2:1mut}.

We can consider only one side of hole as a terminal. Other terminal is part of right border of the plate. The current distribution for this case is shown in Fig. \ref{ht2:side_term}. The inductance of this current path is $3.27pH$. Mutual inductance between current around hole and terminal to terminal current path is $0.084 pH$.

\subsection{Hole as current source}

\begin{figure}
\centering
\subfloat[]
{
\includegraphics[width=0.5\textwidth] {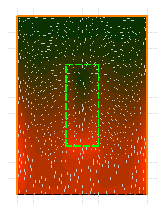}
\label{hv:3v}
}
\subfloat[]
{
\includegraphics[width=0.5\textwidth] {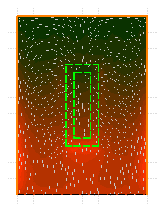}
\label{hv:3cv}
}
\caption{Spatial distribution of supercurrent in $8\times 11\mu m$ plate with $2\times 5\mu m$ hole for different ways to set the current in the plate via internal terminal. Calculations have done  making use of second internal terminal model. 
a) Current is injected in internal area of hole (dashed).
b) Current is injected between coil boundaries (dashed). 
In both cases current leave plate across bottom boundary. Darker gray corresponds higher current density. Current direction is shown by vanes.}
\label{hv:htv}
\end{figure}

Next calculations are performed for model of hole as current source. In this case, area of contact isn't cutted out and there are no hole in the film but current source with homogenous density is present. We consider the same $8\times 11\mu m$ plate with $2\times 5\mu m$ area of current source.

Current directions are shown in Fig. \ref{hv:3v}, the inductance of this current path is $1.61pH$.

The current source can be more complicated. On Fig. \ref{hv:3cv} we demonstrate coil-like source simulating London current crowding effect in the vertical contact. The inductance in this case is $1.58pH$.


\subsection{Multilayer interferometer}

\begin{figure}
\centering
\subfloat[]
{
\includegraphics[width=1.0\textwidth] {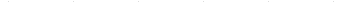}
\label{intf:mesh}
} \newline
\subfloat[]
{
\includegraphics[width=1.0\textwidth] {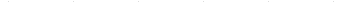}
\label{intf:gp}
} \newline
\subfloat[]
{
\includegraphics[width=1.0\textwidth] {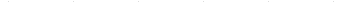}
\label{intf:m2}
} \newline
\subfloat[]
{
\includegraphics[width=1.0\textwidth] {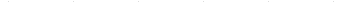}
\label{intf:m3}
}
\caption{ \small {
The layout of interferometer designed for IPHT RSFQ process \cite{CF_IPHT}. Grid size is $10\mu m$. Dark gray corresponds higher current density. Current direction is shown by vanes. \\
a) The result of layout meshing using program "Triangle".
b) Spatial distribution of supercurrent in the groundplane (first metal layer). Groundplane length is $170\mu m$, width is $27.5 \mu m$.
c) Spatial distribution of supercurrent in left net in second layer. Squared dashed internal contact is $10\times 9.5\mu m$
d) Spatial distribution of supercurrent in third layer (metal 3). Width of strip part is $10\mu m$, length is $110\mu m$.
} }
\label{intf:intfb}
\end{figure}

The next example is multilayer interferometer designed for IPHT RSFQ process \cite{CF_IPHT}. Design contains three layers, $0.2\mu m$, $0.25\mu m$ and $0.35\mu m$ thick and  $\lambda = 0.9\mu m.$ The distances between the layers are $0.25\mu m$ and $0.37\mu m$. The shape and dimensions of nets are presented in Fig. \ref{intf:intfb}.

We consider a current circulating in all three layers. The first layer is the ground plane, see Fig. \ref{intf:gp}. The second layer consists of two square parts, see Fig. \ref{intf:m2}. The parts are symmetric so only left part is shown on Fig. \ref{intf:m2}. Ground plane is connected with second layer by two contacts shown on Fig. \ref{intf:gp} and Fig. \ref{intf:m2} by dashed segments.
The second layer, see Fig. \ref{intf:m2}, is connected with the third layer using the square current source terminals shown on Fig. \ref{intf:m2}, Fig. \ref{intf:m3} by dashed squares.

It is assumed that a uniformly distributed supercurrent is injected into the second layer through the dashed segment located in the left part of the second layer. Then it flows across the left rectangle to square dashed current source and jump to the third layer. For third layer inlet current source is left dashed square and outlet source is right dashed square. Then current symmetrically returns back across right part of second layer. First layer carry all return current.

We calculate inductance for this closed current loop. It is $12.6 pH$. The inductance of strip of length $110\mu m$ in third layer over groundplane is $11pH$. 


In Fig. \ref{intf:mesh} mesh of triangles is shown. This mesh is created for all nets once taking into account all projections on bottom layer plane. In this case, the mesh accurately retrace all boundaries of nets. This adaptive non-regular mesh improves accuracy of FEM.



\section{Conclusions}

We developed the new version of 3D-MLSI software for calculation of inductances and currents in complex multilayer superconductor structures. It provides higher accuracy and computing performance.
We have now the simulation tool that allow calculation of inductances, currents and fields of practically all superconductor structures. 
%
%
We significantly improve the numerical algorithm of 3D-MLSI software.  To do this we further developed the
meshing procedure by using well recognized triangulation engine  tool "Triangle" as the core of meshing. Moreover we implemented triangulation for joint projection of all nets thus improving discrete physical model for inductance extraction in layouts designed for planarized and non-planarized fabrication processes.
We introduced two physical model for description of the internal terminals that are staggered or stacked vias between layers or connections between the films contained  Josephson junction. Using simple examples, we have demonstrated their ability to describe the spatial distribution of the currents and to calculate the inductances  in structures and devices having contacts between the individual layers in multilayer designs.

We would like to thank V.K. Semenov, C.J. Fourie, E.B. Goldobin and L.R. Tagirov for fruitful discussions and C.J. Fourie for practical testcase.
M.K. acknowledge partial support by the Program of Competitive Growth of Kazan Federal University.

\section*{References}


\bibliography{ht}

\begin{thebibliography}{}

\bibitem[IAR, 2014]{IARPA}
 (2014).
\newblock Cryogenic computing complexity (\uppercase{C3}):
  http://www.iarpa.gov/index.php/research-programs/c3.

\bibitem[Anders et~al., 2010]{Anders}
Anders, S., Blamire, M., Buchholz, F.-I., Crété, D.-G., Cristiano, R.,
  Febvre, P., Fritzsch, L., Herr, A., Il'Ichev, E., Kohlmann, J., Kunert, J.,
  Meyer, H.-G., Niemeyer, J., Ortlepp, T., Rogalla, H., Schurig, T., Siegel,
  M., Stolz, R., Tarte, E., Ter~Brake, H., Toepfer, H., Villegier, J.-C.,
  Zagoskin, A., and Zorin, A. (2010).
\newblock European roadmap on superconductive electronics - status and
  perspectives.
\newblock {\em Physica C: Superconductivity and its Applications},
  470(23-24):2079--2126.

\bibitem[Bunyk and Rylov, 1993]{Lmeter}
Bunyk, P. and Rylov, S. (1993).
\newblock Automated calculation of mutual inductance matrices of multilayer
  superconductor integrated circuits.
\newblock In {\em Proc. Ext. Abstracts 4th Int. Supercond. Electron. Conf.
  (ISEC�93), Boulder, CO}, page~62.

\bibitem[Duzer and Turner, 1999]{van1998principles}
Duzer, T.~V. and Turner, C.~W. (1999).
\newblock {\em Principles of Superconductive Devices and Circuits, (Second
  Ed.)}.
\newblock Prentice Hall PTR, Upper Saddle River, NJ, USA.

\bibitem[Fourie, 2013]{InductExCalibration}
Fourie, C. (2013).
\newblock Calibration of inductance calculations to measurement data for
  superconductive integrated circuit processes.
\newblock {\em Applied Superconductivity, IEEE Transactions on},
  23(3):1301305--1301305.

\bibitem[Fourie and Volkmann, 2013]{Fourie2013Status}
Fourie, C. and Volkmann, M. (2013).
\newblock Status of superconductor electronic circuit design software.
\newblock {\em Applied Superconductivity, IEEE Transactions on},
  23(3):1300205--1300205.

\bibitem[Fourie et~al., 2013]{CF_IPHT}
Fourie, C., Wetzstein, O., Kunert, J., Toepfer, H., and Meyer, H.-G. (2013).
\newblock Experimentally verified inductance extraction and parameter study for
  superconductive integrated circuit wires crossing ground plane holes.
\newblock {\em Superconductor Science and Technology}, 26(1):015016.

\bibitem[Fourie et~al., 2011]{CoenradFourier2011}
Fourie, C., Wetzstein, O., Ortlepp, T., and Kunert, J. (2011).
\newblock Three-dimensional multi-terminal superconductive integrated circuit
  inductance extraction.
\newblock {\em Superconductor Science and Technology}, 24(12):125015.

\bibitem[Fourier, 2014]{InducEx}
Fourier, C. (2014).
\newblock Inductex version 4.26. online:
  http://stbweb02.stb.sun.ac.za/inductex/.

\bibitem[Fujimaki et~al., 2014]{Fujima}
Fujimaki, A., Tanaka, M., Kasagi, R., Takagi, K., Okada, M., Hayakawa, Y.,
  Takata, K., Akaike, H., Yoshikawa, N., Nagasawa, S., Takagi, K., and Takagi,
  N. (2014).
\newblock Large-scale integrated circuit design based on a \uppercase{N}b
  nine-layer structure for reconfigurable data-path processors.
\newblock {\em IEICE Transactions on Electronics}, E97-C(3):157--165.

\bibitem[Gaj et~al., 1999]{GajFeldman}
Gaj, K., Herr, Q., Adler, V., Krasniewski, A., Friedman, E., and Feldman, M.
  (1999).
\newblock Tools for the computer-aided design of multigigahertz superconducting
  digital circuits.
\newblock {\em Applied Superconductivity, IEEE Transactions on}, 9(1):18--38.

\bibitem[Jin, 2002]{jin2002finite}
Jin, J. (2002).
\newblock {\em The Finite Element Method in Electromagnetics}.
\newblock A Wiley-Interscience publication. Wiley.

\bibitem[Kamon~M. and J.K., 1994]{FASHENRY0}
Kamon~M., T.~M. and J.K., W. (1994).
\newblock Fasthenry: a multipole-accelerated 3-\uppercase{D} inductance
  extraction program.
\newblock {\em Microwave Theory and Techniques, IEEE Transactions on},
  42(9):1750--1758.

\bibitem[Khapaev, 1996]{KhapaevJr}
Khapaev, M. (1996).
\newblock Extraction of inductances of a multi-superconductor transmission
  line.
\newblock {\em Superconductor Science and Technology}, 9(9):729--733.

\bibitem[Khapaev, 2001]{khapaev2001inductance}
Khapaev, M. (2001).
\newblock Inductance extraction of multilayer finite-thickness superconductor
  circuits.
\newblock {\em Microwave Theory and Techniques, IEEE Transactions on},
  49(1):217--220.

\bibitem[Khapaev et~al., 2001]{khapaev20013d}
Khapaev, M., Kidiyarova-Shevchenko, A., Magnelind, P., and Kupriyanov, M.
  (2001).
\newblock \uppercase{3D-MLSI}: software package for inductance calculation in
  multilayer superconducting integrated circuits.
\newblock {\em Applied Superconductivity, IEEE Transactions on},
  11(1):1090--1093.

\bibitem[Khapaev and Kupriyanov, 2010a]{kupriyanov2010}
Khapaev, M. and Kupriyanov, M. (2010a).
\newblock Sheet current model for inductances extraction and josephson
  junctions devices simulation.
\newblock {\em Journal of Physics: Conference Series},
  248:012037(1)--012037(8).

\bibitem[Khapaev and Kupriyanov, 2010b]{TEEconf}
Khapaev, M. and Kupriyanov, M. (2010b).
\newblock {\em Sparse Approximation of FEM Matrix for Sheet Current
  Integro-Differential Equation}, chapter~33, pages 510--522.
\newblock World Scientific.

\bibitem[Khapaev et~al., 2003]{meSST}
Khapaev, M., Kupriyanov, M., Goldobin, E., and Siegel, M. (2003).
\newblock Current distribution simulation for superconducting multi-layered
  structures.
\newblock {\em Superconductor Science and Technology}, 16(1):24.

\bibitem[Nagasawa et~al., 2014]{Nagasawa2}
Nagasawa, S., Hinode, K., Satoh, T., Hidaka, M., Akaike, H., Fujimaki, A.,
  Yoshikawa, N., Takagi, K., and Takagi, N. (2014).
\newblock \uppercase{Nb} 9-layer fabrication process for superconducting
  large-scale \uppercase{SFQ} circuits and its process evaluation.
\newblock {\em IEICE Transactions on Electronics}, E97-C(3):132--140.

\bibitem[Nagasawa et~al., 2009]{Nagasawa}
Nagasawa, S., Satoh, T., Hinode, K., Kitagawa, Y., Hidaka, M., Akaike, H.,
  Fujimaki, A., Takagi, K., Takagi, N., and Yoshikawa, N. (2009).
\newblock New \uppercase{N}b multi-layer fabrication process for large-scale
  \uppercase{SFQ} circuits.
\newblock {\em Physica C: Superconductivity and its Applications},
  469(15-20):1578--1584.

\bibitem[Orlando and Delin, 1991]{Orlando:book}
Orlando, T. and Delin, K. (1991).
\newblock {\em Foundations of Applied Superconductivity}.
\newblock Electrical Engineering Series. Addison-Wesley.

\bibitem[Ren, 2003]{ZRen2003}
Ren, Z. (2003).
\newblock 2-\uppercase{D} dual finite-element formulations for the fast
  extraction of circuit parameters in \uppercase{VLSI}.
\newblock {\em Magnetics, IEEE Transactions on}, 39(3):1590--1593.

\bibitem[Rubinacci and Tamburrino, 2010]{Tamburrino2010}
Rubinacci, G. and Tamburrino, A. (2010).
\newblock Automatic treatment of multiply connected regions in integral
  formulations.
\newblock {\em Magnetics, IEEE Transactions on}, 46(8):2791--2794.

\bibitem[Ruehli, 1974]{peecRuehli}
Ruehli, A. (1974).
\newblock Equivalent circuit models for three-dimensional multiconductor
  systems.
\newblock {\em Microwave Theory and Techniques, IEEE Transactions on},
  22(3):216--221.

\bibitem[Shewchuk, 1996]{shewchuk96b}
Shewchuk, J. (1996).
\newblock Triangle: {E}ngineering a {2D} {Q}uality {M}esh {G}enerator and
  {D}elaunay {T}riangulator.
\newblock In Lin, M.~C. and Manocha, D., editors, {\em Applied Computational
  Geometry: Towards Geometric Engineering}, volume 1148 of {\em Lecture Notes
  in Computer Science}, pages 203--222. Springer-Verlag.
\newblock From the First ACM Workshop on Applied Computational Geometry.

\bibitem[Tolpygo et~al., 2014a]{Tolpygo22014}
Tolpygo, S., Bolkhovsky, V., Weir, T., Johnson, L., Oliver, W., and Gouker, M.
  (2014a).
\newblock Deep sub-micron stud-via technology for superconductor
  \uppercase{VLSI} circuits.
\newblock {\em Journal of Physics: Conference Series}, 507(PART 4).

\bibitem[Tolpygo et~al., 2014b]{Tolpygo2014}
Tolpygo, S., Bolkhovsky, V., Weir, T., Johnson, L., Oliver, W., and Gouker, M.
  (2014b).
\newblock Deep sub-micron stud-via technology of superconductor
  \uppercase{VLSI} circuits.
\newblock {\em Superconductor Science and Technology}, 27(2).

\bibitem[{Tolpygo} et~al., 2014]{2014arXiv1408.5828T}
{Tolpygo}, S.~K., {Bolkhovsky}, V., {Weir}, T.~J., {Galbraith}, C.~J.,
  {Johnson}, L.~M., {Gouker}, M.~A., and {Semenov}, V.~K. (2014).
\newblock {Inductance of Circuit Structures for MIT LL Superconductor
  Electronics Fabrication Process with 8 Niobium Layers}.
\newblock {\em ArXiv e-prints}.

\bibitem[Whiteley, 2014]{FastHenry3.0wr}
Whiteley, S.~R. (2014).
\newblock Fasthenry 3.0wr: http://www.wrcad.com/ftp/pub/.

\end{thebibliography}


\end{document}